\documentclass[11pt]{article}

\usepackage{amsthm}
\usepackage{graphicx}
\usepackage{epstopdf}
\usepackage{amsmath,amstext,amssymb,amsfonts,verbatim}
\usepackage{xcolor}
\usepackage{hyperref}
\usepackage{algorithmic,algorithm,rotating}

\hypersetup{colorlinks=false,linkbordercolor=red,linkcolor=blue,pdfborderstyle={/S/U/W 1}}

\theoremstyle{definition}

\title{\vspace{-2.5cm}Algorithmic Complexity and Reprogrammability of Chemical Structure Networks\thanks{An online implementation to estimations of graph complexity is available online at \url{http://www.complexitycalculator.com}}}

\author{Hector Zenil$^{1,2,3,4}$, Narsis A. Kiani$^{1,2,3,4}$, Ming-Mei Shang $^{2,3}$ \\and Jesper Tegn\'er$^{2,3,5}$\\
$^1$ Algorithmic Dynamics Lab, Centre for Molecular Medicine,\\ Karolinska Institute, Stockholm, Sweden\\
$^2$ Unit of Computational Medicine, Department of Medicine,\\Karolinska Institute, Stockholm, Sweden \\
$^3$ Science for Life Laboratory, SciLifeLab, Stockholm, Sweden\\
$^4$ Algorithmic Nature	Group, LABORES	for	the	Natural	and\\Digital Sciences, Paris, France\\
$^5$ Biological and Environmental Sciences and Engineering Division,\\Computer, Electrical and Mathematical Sciences and Engineering\\Division, King Abdullah University of Science and\\	Technology (KAUST), Kingdom	of Saudi Arabia\\
{\{hector.zenil, jesper.tegner\}@ki.se}}

\date{}

\begin{document}

\maketitle

\begin{abstract}
Here we address the challenge of profiling causal properties and tracking the transformation of chemical compounds from an algorithmic perspective. We explore the potential of applying a computational interventional calculus based on the principles of algorithmic probability to chemical structure networks. We profile the sensitivity of the elements and covalent bonds in a chemical structure network algorithmically, asking whether reprogrammability affords information about thermodynamic and chemical processes involved in the transformation of different compound classes. We arrive at numerical results suggesting a correspondence between some physical, structural and functional properties. Our methods are capable of separating chemical classes that reflect functional and natural differences without considering any information about atomic and molecular properties. We conclude that these methods, with their links to chemoinformatics via algorithmic, probability hold promise for future research. \\
    
\end{abstract}

\section{Background and Preliminaries}

One of the major challenges in modern physics is to provide proper and suitable representations of network systems for use in fields ranging from physics~\cite{boccaletti} to chemistry~\cite{chen2014entropy}. A common problem is the description of order parameters with which to characterize the `\textit{complexity of a network}'. Graph complexity has traditionally been characterized using graph-theoretic measures such as degree distribution, clustering coefficient, edge density, and community or modular structure.

A previous algorithmic information-theoretic view of systems toxicity applicable both to network analysis and pharmacokinetic analysis has been proposed ~\cite{toxicologypaper}, with the overarching aim of not only describing but also seeking out causal mechanisms. The suggestion was that since the problem of designing new compounds, aiming to develop drugs, for new targets is challenging, whereas the prediction problem is easier from an inference point-of-view compared to elucidating the mechanisms driving toxicity, complementary approaches are warranted.

For example, instead of engineering a drug to target a unique pathway or mutation of a tiny subset of diseases, drug repositioning involves starting with approved drugs to find combinations that can be used to treat diseases other than the ones they were designed for, with the advantage that approved drugs can bypass much regulation if we correctly control for the effects they can have. Thus prediction and simulation are key. This means that the whole field has to move towards causal modelling and functional inference rather than employing traditional statistical and purely geometric approaches (e.g. distances between compounds or grid-based docking).

Here we are interested in combining techniques originating in fundamental mathematics and theoretical computer science to take a fresh look at long-standing challenges in molecular complexity from an algorithmic information perspective as applied to networks~\cite{zenilkianitegner,postchina}.

Algorithmic information indices may facilitate the characterization of some properties of chemical compounds. Statins, for example, are associated with the heart and cholesterol, while morphine, codeine and heroin share structural properties and effects. Algorithmic information-theoretic approaches like the one showcased here are concerned with predictive causal models, going beyond statistical/descriptive approaches (such as structural alignments). This is important because, for statins, for example, block the cholesterol synthesis pathway by inhibiting the HMG-CoA reductase because similarity to HMG-CoA structure, which is the rationale used to treat cardiovascular diseases in patients. The algorithmic approach deployed here is thus equipped to find model-based mechanistic candidates for this kind of causal processes involved in the transformation and interactions of compounds. Here we will focus on one kind of representation namely chemical structure networks, first independent of physical properties that can then be tested and connected back to find what is a consequence of the compound causal structure/topology and what is a consequence of other intrinsic features such as atomic charge and thermodynamic constraints.

For this approach we will use the techniques and methods developed in ~\cite{zenilkianitegner,maininfo}. The basic idea is to estimate the likelihood of similarity between compounds based on an induced partition of possible common underlying mechanisms (models are found by an exhaustive algorithm that explains small pieces of the data). This is, in general, a hard if not impossible task (uncomputable), but approximations have been shown to be useful and new numerical methods have been advanced that are complementary to previous approaches such as the use of lossless compression algorithms to approximate algorithmic complexity, which are very limited at accounting for causation~\cite{bdm,emergenceofuniversaldistribuion}. Moreover, even when the method is based on the idea of finding minimal programs, the more practical aim is to find any, or a set of programs, explaining the data rather than the smallest one, and so the problem becomes computationally feasible~\cite{d4,d5,kolmo2d,bdm}.

\subsection{Chemical notation}

There are two main notations for chemical substances. The simplified molecular-input line-entry system or SMILES is an ASCII string specification describing the structure of a chemical. The string is obtained by printing the symbol nodes encountered in a depth-first tree traversal of the chemical graph. The chemical graph is first trimmed to remove hydrogen atoms, and cycles are broken to turn it into a spanning tree. Where cycles have been broken, numerical suffix labels are included to indicate the connected nodes, and parentheses are used to indicate points of branching on the tree. For example, nicotine is written as CN1CCC[C@H]1c2cccnc2. A SMILES string thus encodes and contains information about a molecule and is an upper bound of its information content.

A more standardized notation in chemistry is the IUPAC International Chemical Identifier or InChI, another textual identifier for chemical substances, its chief advantage over SMILES being that the InChI algorithm converts the structural information of the chemical substance in a 3-step process that can be tweaked to a desired level of structural chemical detail, except for an unchanged substring representing the substance (called the main layer). The algorithm then removes redundant information, keeps as much of the information of the structure as desired, and encodes it in a string. In some sense, InChI is thus a tighter bound of the algorithmic complexity of the chemical structure captured by ASCII strings and an improvement over SMILES.

\subsection{Molecular complexity}

The concept of molecular complexity has been shown to be relevant in the design of syntheses through minimizing the sum of molecular complexities of the synthetic intermediates~\cite{hendrickson}. A number of proposals have been made in the literature for defining molecular complexity. For example, enumerations of graph invariants for comparing chemical structures have been used for at least 3 decades, and QSAR regression models~\cite{nantasenamat} for even longer. Different approximations emphasize different aspects of the molecule and are heavily observer dependent, because the observer has to make a pre-selection of features of interest (e.g. clustering coefficient, some eigenvalues in graph spectra, degree distributions, etc.).  This is what the chemistry community has been doing with what they call  ``chemical fingerprints'', and prior to that in molecular formulae comparing degree distributions. 

In 1981, Bertz~\cite{bertz} introduced a measure of molecular complexity by applying Shannon's entropy to the distribution of subgraphs in molecular graphs. That was the starting point of a systematic search in chemical theory for relevant measures of molecular complexity. Ever since, graph and molecular complexity measures have focused on statistics of the topological properties of graphs, such as the size of the non-repeating subgraph set, among similar approaches. For recent results and a survey see ~\cite{dehmer, mowshowitz}, including a proposal for using lossless compression as a graph complexity index~\cite{peshkin}. An up-to-date review of mainstream techniques in the area of molecular networks can be found in~\cite{csermely}.

Molecular complexity is not easy to define or to quantify, and all previous approaches have focused on combinatorial or statistical properties of the molecular graphs, either as a function of bond connectivities, specificity of structures or diversity of elements. Researchers agree that the complexity of a molecule increases with increasing size, increasing branching, and increasing cyclicity for acyclic and cyclic structures~\cite{rucker}, but no computable measure can cover all possible enumerable computable features like these (both currently defined and undefined) at the same time. Indeed, a more universal and robust measure of molecular complexity should take into account all these features of interest at the same time, without having to enumerate them explicitly or to define an ad-hoc measure for each of them. A measure that only focuses on some of these properties in the expectation that it could later be generalized to be able to deal with other properties is out of the question. 

Small Molecules or compounds are commonly represented by their skeletal molecular graphs (see Fig.~\ref{a}A). That is, the union of a set of points, symbolizing atoms other than hydrogen, and a set of lines, symbolizing molecular bonds. A typical similarity formula (see Fig.~\ref{a}E-F) is given by the total number of elements in the bin 0 divided by the square of the total count of atoms of the largest molecule. The closer to 0, the more dissimilar. The formula can be relaxed by taking near 0 bin elements, but this only works for the most simple cases of structural similarity. This kind of approach is descriptive rather than predictive. For example, regardless of different biological mechanisms of action, aspirin and statins have shown similar beneficial effects on cardiovascular diseases (CVDs) at population level, combination usage of the two drugs has additive effects, but it is not yet clear in what precise ways aspirin and statins differ, while the general agreement is that they possess similarities (e.g. accumulating evidence from basic and observational research demonstrate the anti-inflammatory effects of both drugs contribute to CVDs treatment, and combined usage of two drugs has additive and synergistic effects over the use of only one\cite{platelets}). It is not difficult to see how in some cases compounds or compound substructures may share information from complementary regions, e.g. between drugs and targets, because the structure of the docking entity is energetically and structurally the complement of the docking region of the other entity, and thus the compounds may display (partially or entirely) similar classical and algorithmic information properties and estimations. 

Molecular complexity is not easy to define or to quantify, and all previous approaches have focused on combinatorial or statistical properties of the molecular graphs, either as a function of bond connectivities, specificity of structures or diversity of elements. Researchers agree that the complexity of a molecule increases with increasing size, increasing branching, and increasing cyclicity for acyclic and cyclic structures~\cite{rucker}, but no computable measure can cover all possible enumerable computable features like these (both currently defined and undefined) at the same time. Indeed, a more universal and robust measure of molecular complexity should take into account all these features of interest at the same time, without having to enumerate them explicitly or to define an ad-hoc measure for each of them. 

The number of possible statistical and algorithmic properties in all possible networks is countably infinite, but no effective (computable) measure can account for all of them~\cite{martinlof}. This only leave us with uncomputable measures that can serve as general universal measures of complexity equipped to find any effective (statistical or algorithmic) regularity. Our approach may find some applications. For example, one may find that \textit{low algorithmic complexity} molecules are easier to synthesize or to assemble into larger new molecules and drugs because \textit{high algorithmic complexity} molecules would share fewer physical properties.

\subsection{Causality and Algorithmic Probability}

The concept of algorithmic complexity~\cite{kolmo,chaitin} is at the core of the challenge of complexity in discrete dynamic systems, as it involves finding the most statistically likely generating mechanism (computer program) that produces some given data. Formally, the algorithmic complexity (also known as Kolmogorov-Chaitin complexity) is the length of the shortest computer program that reproduces the data from its compressed form when running on a universal Turing machine.

We follow the so-called Coding Theorem (CTM) and Block Decomposition Methods (BDM) as introduced in~\cite{d4,d5,kolmo2d,bdm}, based on the seminal concept of Algorithmic Probability~\cite{solomonoff,levin}, which in turn is strongly related to algorithmic complexity~\cite{kolmo,chaitin}. The only parameters used for the decomposition of BDM as suggested in~\cite{bdm} was the maximum 12 for strings and 4 for arrays given the current best CTM approximation~\cite{d5} based on an empirical distribution based on all Turing machines with up to 5 states, and no string/array overlapping decomposition for maximum efficiency (as it runs in linear time) and for which the error (due to boundary conditions) is bounded~\cite{bdm}. However, the algorithm introduced here is independent of the method used to approximate algorithmic complexity, such as BDM. BDM assigns an index associated with the size of the most likely generating mechanism producing the data according to Algorithmic Probability~\cite{solomonoff}. BDM is capable of capturing features in data beyond statistical properties~\cite{bdm,zkgraph}, and thus represents an improvement over classical information theory. Because finding the program that reproduces a large object is computationally very expensive even to approximate, BDM finds short candidate programs (which are generative models) using a method introduced in~\cite{d4,d5} that finds and reproduces fragments of the original object and then puts them together as a candidate algorithmic model of the whole object~\cite{bdm,kolmo2d}. These short computer programs are effectively candidate mechanistic models explaining each fragment, with the long finite sequence of short models being itself a generating mechanism. 

In this sense, a \textit{causal path} is a path where the changes between one state and another is merely the product of an underlying dynamical system following its normal course, sans external intervention~\cite{maininfo}. 

\begin{figure}
\centering

\textbf{A}\hspace{2.5cm}\textbf{B}\hspace{3cm}\textbf{C}\hspace{2.2cm}\textbf{D}\\

\medskip

\scalebox{.3}{\includegraphics{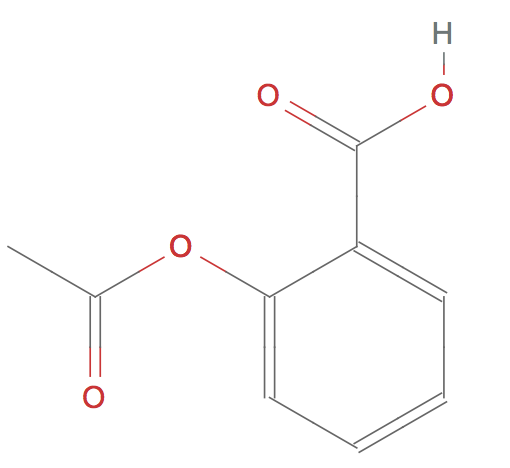}}\hspace{.24cm}
\scalebox{.16}{\includegraphics{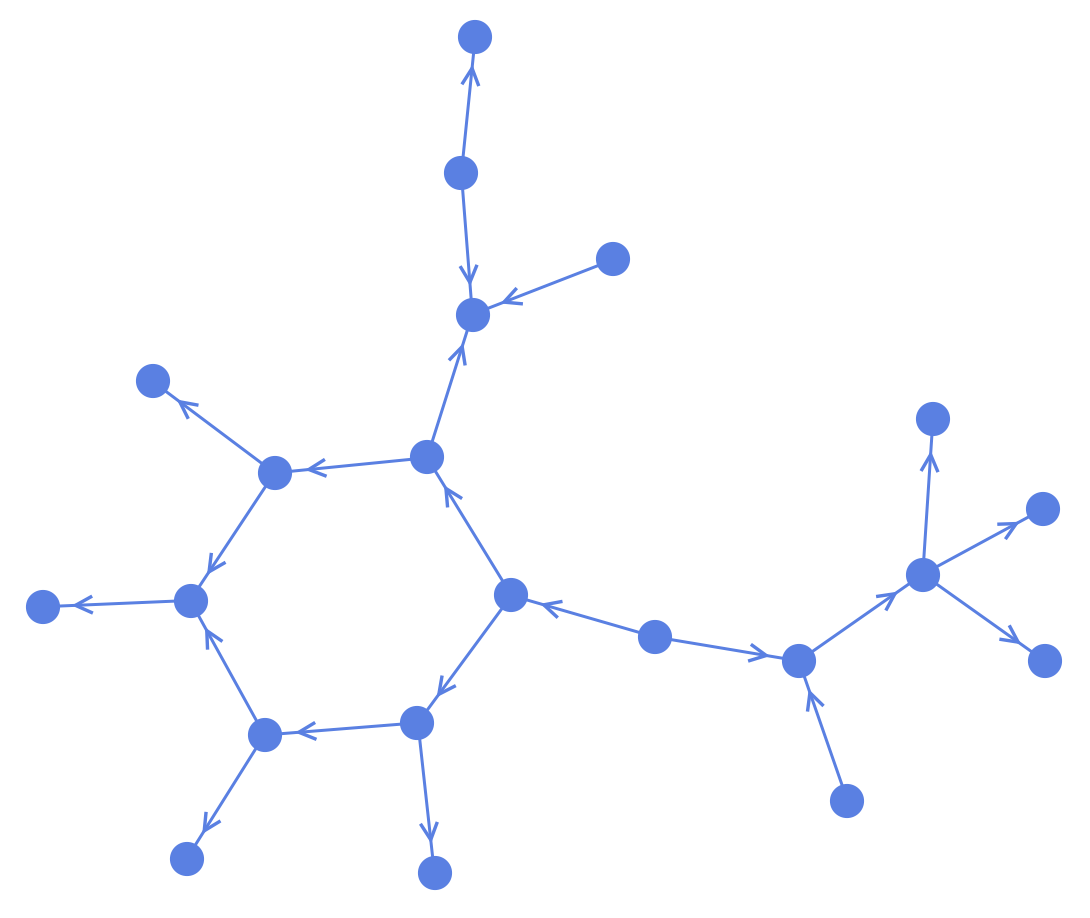}}\hspace{.24cm}
\scalebox{.22}{\includegraphics{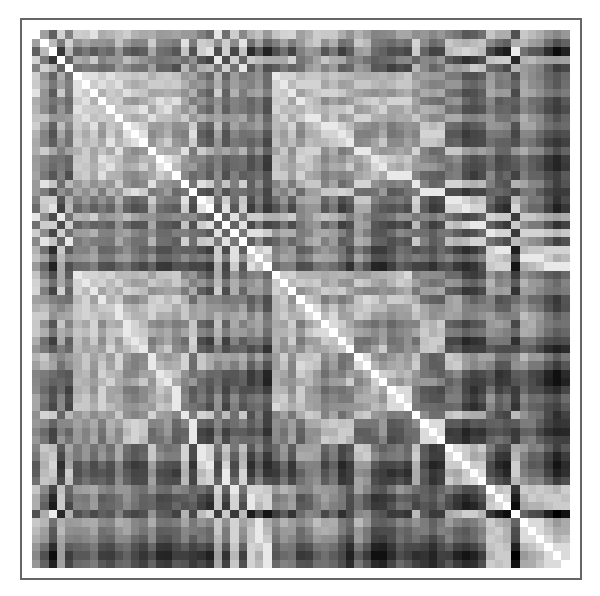}}\hspace{.2cm}
\scalebox{.22}{\includegraphics{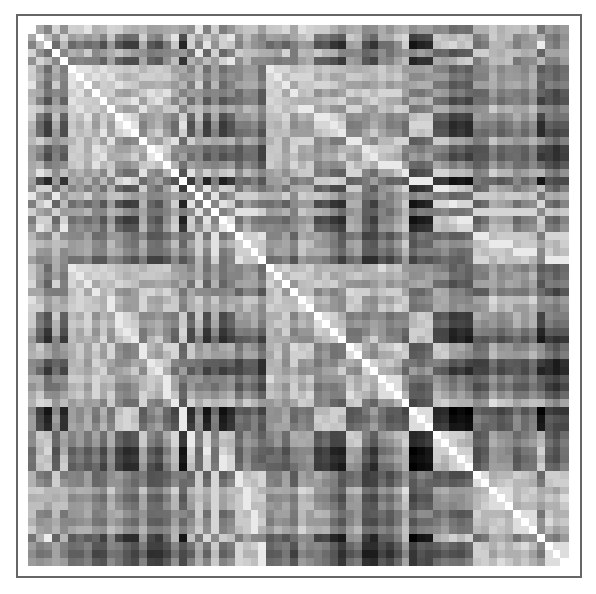}}\\

\medskip

\textbf{E}\hspace{5.2cm}\textbf{F}

\medskip

\scalebox{.24}{\includegraphics{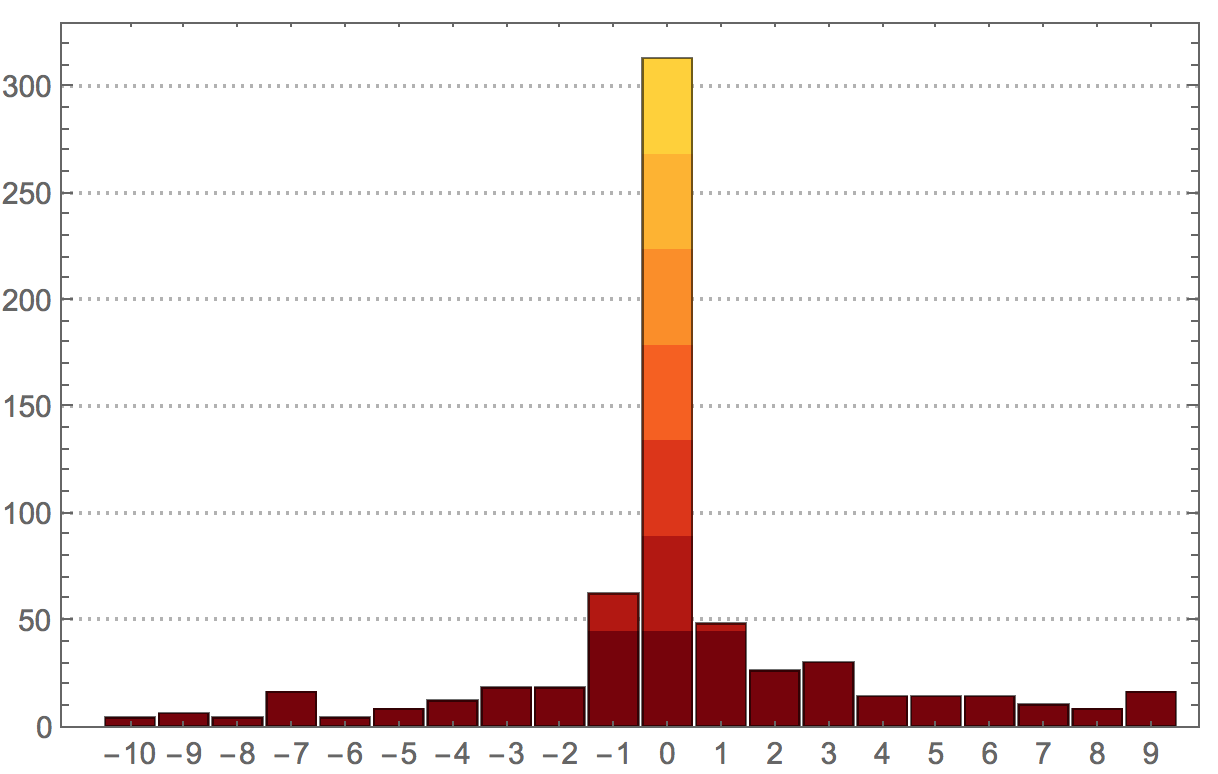}}\hspace{.8cm} \scalebox{.24}{\includegraphics{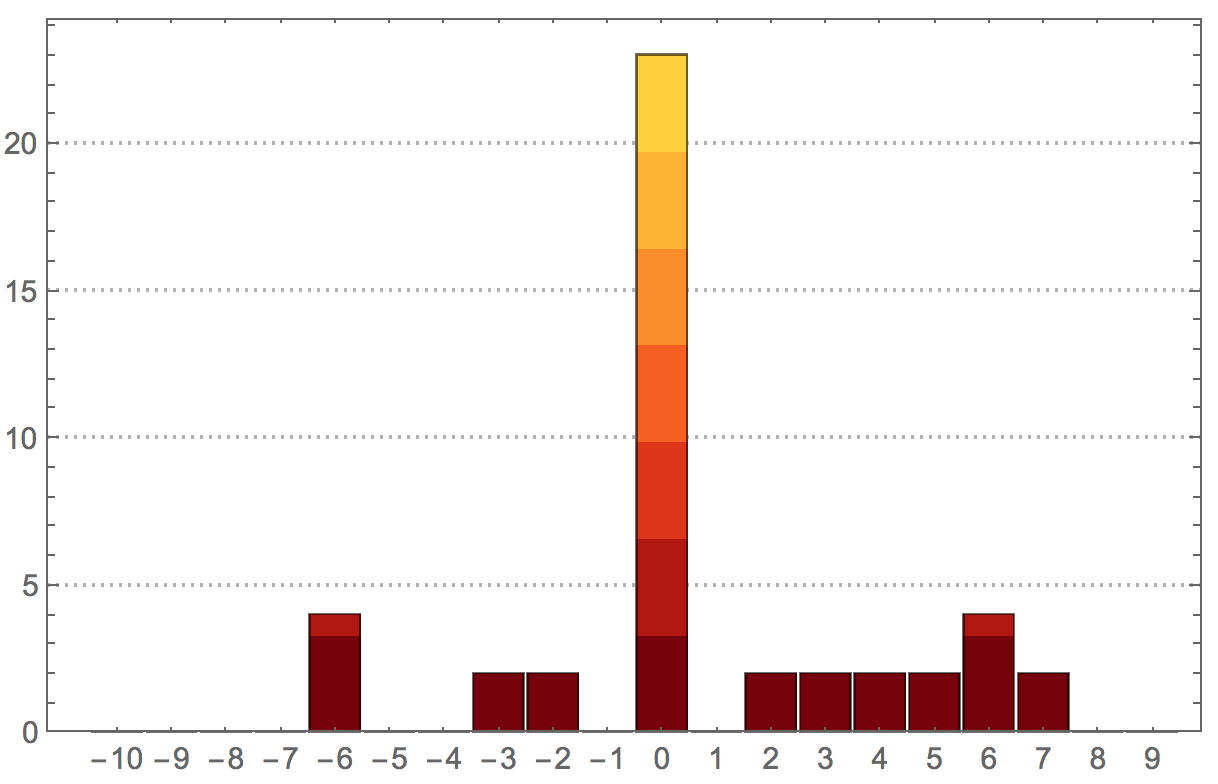}}\\

\medskip

\textbf{G}\hspace{6cm}\textbf{H}

\medskip

\scalebox{.24}{\includegraphics{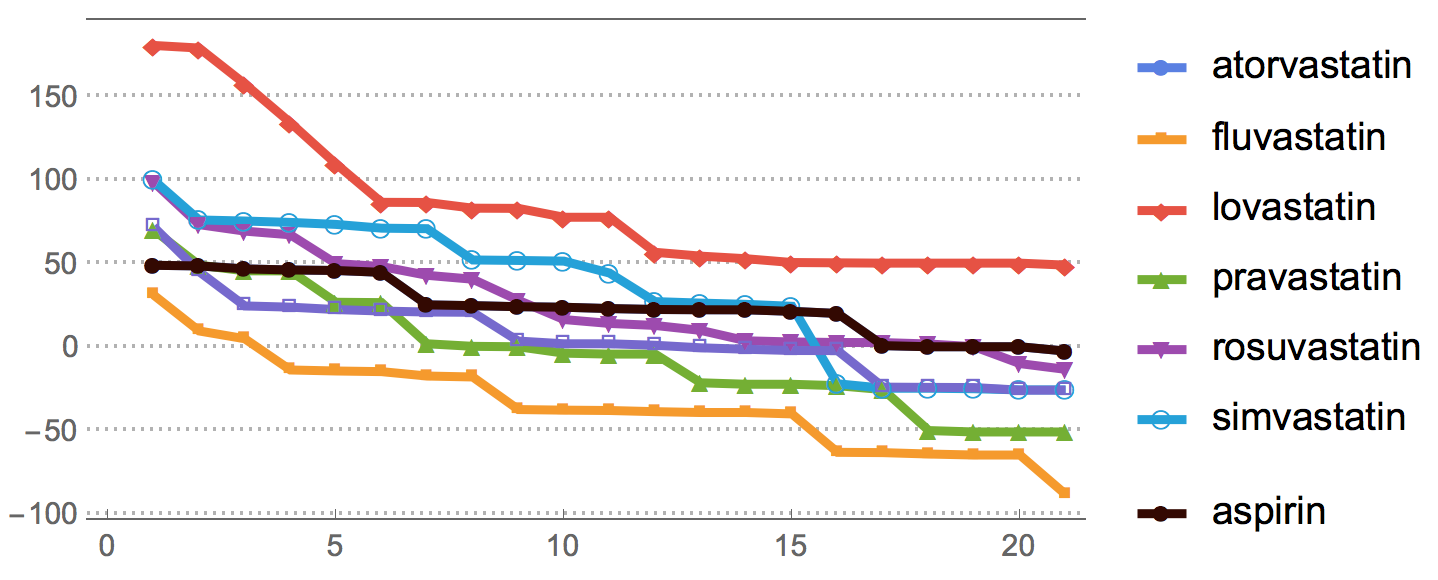}}\hspace{.5cm}\scalebox{.26}{\includegraphics{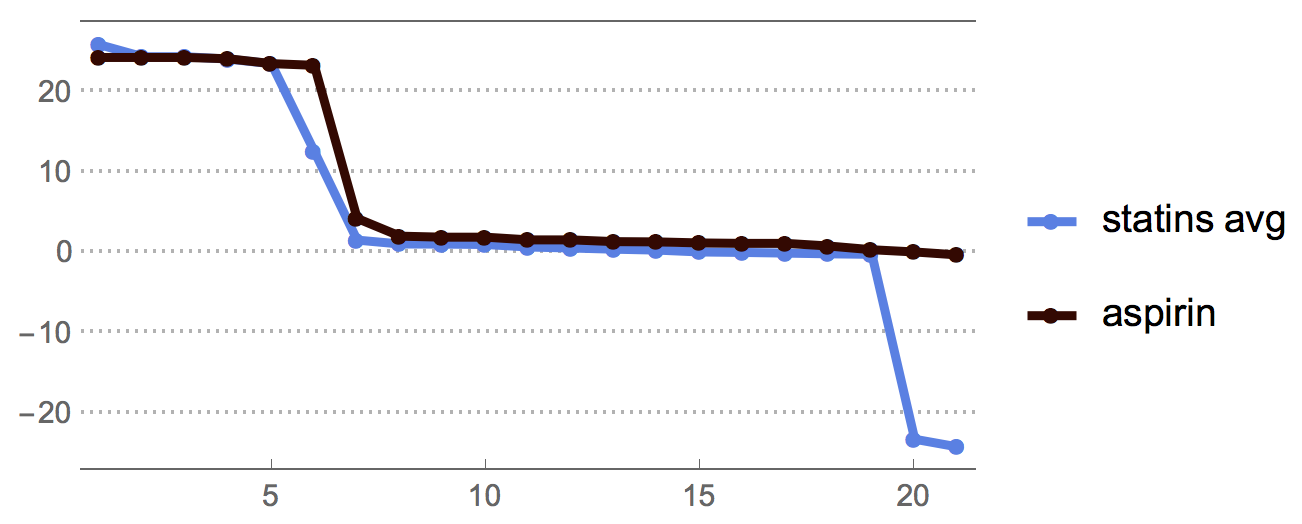}}\\

\caption{\label{a}A: Canonical structural diagram of aspirin. B: The undirected chemical structure network of aspirin, where shape, particular elements and bond types are not retained. C and D: Statins have similar contact maps. Here we feature lovastatin and simvastatin. Contact maps of a compound are calculated from the distance among its constituent atoms; the further away, the darker. E: Alignment histograms between lovastatin and simvastatin. The greater the number of atoms in or around the zero bin, the more similar. The $x$-axis represents the average distance among atoms. F: Weaker alignment between lovastatin and aspirin than among statins. G: Algorithmic (mis)alignment from the node information signatures of aspirin versus statins (normalized by aspirin size) with lovastatin topping all others. H: The edge information signature of aspirin is all positive i.e. all single-bond perturbations to aspirin make its generative mechanistic model even simpler. In comparison, statins (average signature normalized by aspirin size) have very similar signatures but differ in the number of molecular bonds that when removed send the compound networks towards algorithmic randomness.}
\end{figure}

An important concept is that of the \textit{information signature} of an object~\cite{maininfo}. An information signature quantifies the algorithmic resilience of an object to transformations, that is how much its most likely mechanistic model may change after modifying the object. In the case of networks, perturbations can be applied to nodes or edges, that is, in the context of chemical structure networks to atoms and molecular bonds which means that one can have both node and edge information signatures (see Fig.~\ref{a}G,H). Comparing information signatures is therefore a way to perform an algorithmic alignment among different objects such as chemical compounds.

\section{Numerical Experiments and Results}

\subsection{Algorithmic structural complexity} 

We perturb the structure of a chemical compound network and see the effect on the set of candidate generating models by performing interventions and ranking them by the disruptiveness and causal contribution to the networks' original algorithmic information content and therefore to the networks' original hypothesize generative models (as found by the CTM/BDM method).

We may attach the rubric \textit{in silico alchemy} to the digital computer simulation of the types of changes that a molecule can be subject to regardless of the thermodynamic aspects of said molecule or the processes involved (later, we will compare it to the known processes and physical properties associated with the old and new compounds). Central to the ideas exploited here is the notion of the information signature as the result of an in silico simulation measuring the sensitivity of causal generative model of a compound to perturbations. The information signature depicted in Fig.~\ref{a}G illustrates a set of such interventions/perturbations simulating the kinds of transformations that a compound such as aspirin can undergo, measuring the structural sensitivity to single-bond changes and the susceptibility of aspirin to being reprogrammed (artificially converted) into a different or similar (causal) structure in what would be a causal path. 

This is an alchemy of sorts, because some of these transformations may be thermodynamically unlikely and the simulation takes no account of any physical properties (one can easily transform any element into gold under such conditions). But all likely causally topological paths are studied as equally possible in order to determine if there are thermodynamic effects that can be explained by algorithmic causality,
rather than being external effects following particular laws but being intrinsic properties of the compound in question.

\subsection{Chemical network perturbation analysis}

All values in the node (Fig.~\ref{a}G) and edge (Fig.~\ref{a}H) information signatures of aspirin are positive, that is, no intervention targeting any element pushes aspirin to become a more complex compound. Their algorithmic alignment is similar to the one produced by classical geometric alignments yet there are less arbitrary cutoff values (atom to atom distance threshold). The result is consistent with the literature characterizing aspirin as a simple structure. More importantly, the signature of aspirin has two clearly identifiable regions. In the disruptive regime at about 25 bits---measuring the difference between the mutated/disrupted compound and the original aspirin structure---are the elements of the carbon ring that is identified as the most stable structure in aspirin. The reduction in algorithmic complexity comes from the fact that breaking the carbon cycle produces a simple tree graph with a long path graph, a graph that is of even lower algorithmic complexity because there is an even shorter program that can produce a tree with a long path than a structure with the ring/cycle. In contrast, removing all hydrogen and oxygen elements makes an algorithmically neutral contribution, meaning that their removal is less disruptive to the core of the structure of aspirin (even though it may be more deeply implicated in its functioning, a limitation of this type of analysis if, e.g., valency values or electric charges are not incorporated in the network description (e.g. as weights)--- which it is possible to do though we do not cover it in this paper). Yet the signature analysis indicates that such elements may more easily be found in more causal paths than atoms from the carbon ring. In other words, it is algorithmically less random to find a carbon ring in the middle of a structure than to add some elements to the molecule by, e.g., methylation or phosphorylation. Another observation from Fig.~\ref{a}H is that fluvastatin has the most negative node information signature among the statins and it is also the statin with the lowest number of interactions compared to most other statins and lovastatin is the most positive and also similar to aspirin in its information changes remaining algorithmic simple, with all node and edge perturbations positive. Lovastatin has similar interactions to atorvastatin and simvastatin which are also close to Lovastatin in the signature information landscape. Aspirin is, however, in the middle of the statins pack suggesting a greater similarity than what classical alignment methods suggest (Fig.~\ref{a}F).

\subsection{Algorithmic causal transformations}

\begin{figure}
\centering

\textbf{A}\hspace{6cm} \textbf{B}\\

\medskip

\scalebox{.22}{\includegraphics{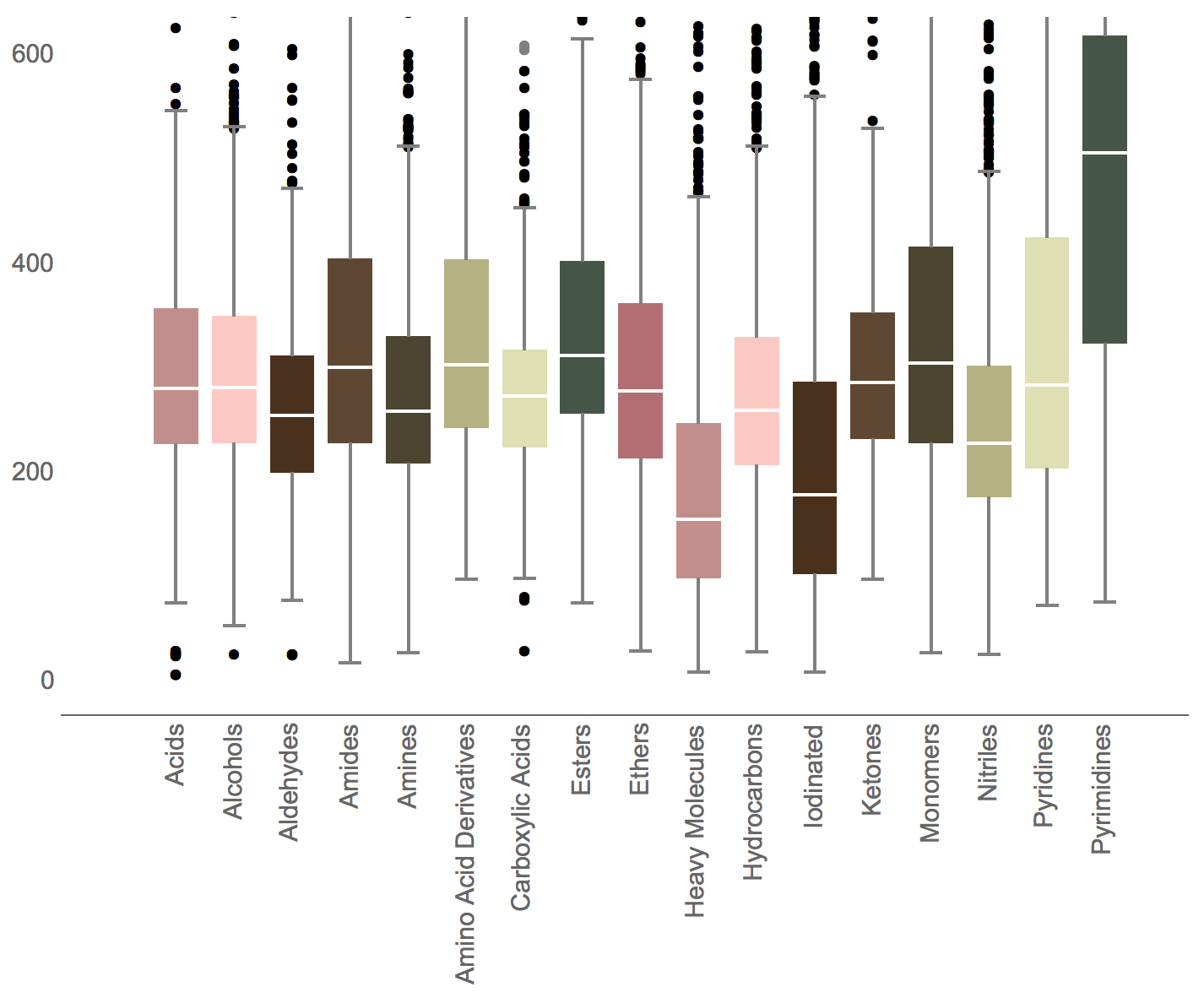}}\scalebox{.26}{\includegraphics{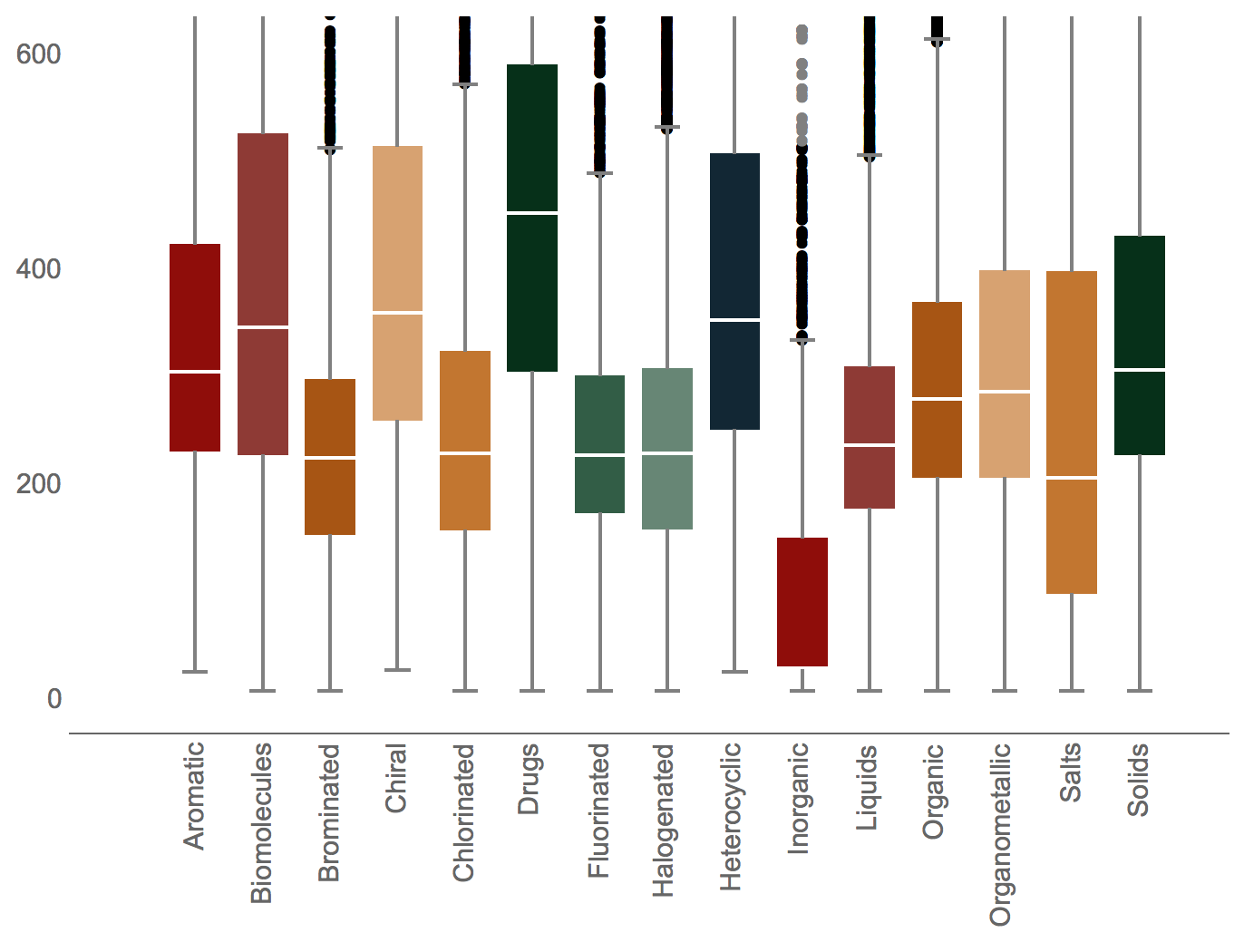}}\\

\textbf{C}\hspace{4.5cm} \textbf{D}\\

\medskip

\hspace{1cm}\scalebox{.365}{\includegraphics{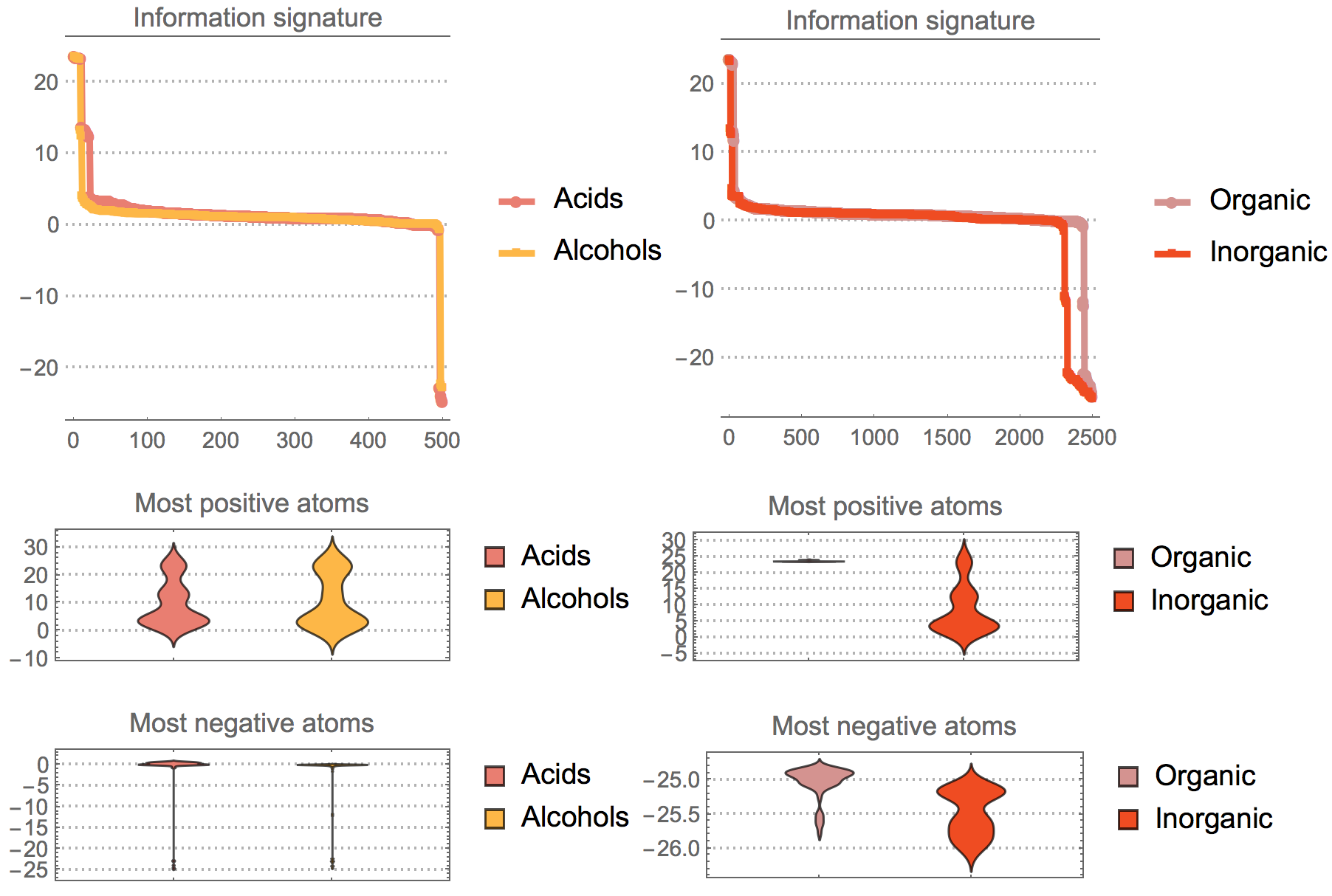}}

\caption{\label{b}A and B: Classification of compound classes according to their estimations of algorithmic probability/complexity by BDM. C and D: In silico alchemy by intervening in chemical structure networks and analyzing their algorithmic properties. The algorithmic sensitivity of acids and  alcohols is very similar, corresponding to known mechanistic processes that can transform one into the other. A minor asymmetry can be found similar to the difficulty of converting one compound into another. In contrast, organic vs inorganic compounds are among the most dissimilar.}
\end{figure}

Chemical compounds that may be in the same causal path will have similar signatures~\cite{maininfo} from single bond or atomic knock-out interventions. One such example may be that of acids and alcohols compared to compounds that are structurally---and causal structurally
more removed---such as compounds one may expect to form between organic versus inorganic substances (as tested in Fig.~\ref{b}D). This can be seen by analyzing the tails of the signature distributions (Fig.~\ref{b}C-D) that provide a network with the means of moving towards or away from its original algorithmic model. 

It is also of interest to find a correspondence between the algorithmic difficulty of transforming a compound such as an acid to an alcohol. What is suggested in the preliminary experiments is that it is (slightly) more difficult because it implies a reduction of algorithmic randomness (Fig.~\ref{b}C), as compared to transforming an alcohol to an acid, which is consistent with the literature showing that oxidants able to perform this operation in, e.g,. complex organic molecules require substantial selectivity, therefore making it less likely to happen by the chance imposition of a thermodynamic direction, something 
also suggested by the algorithmic causal calculus. This is, however, less dramatic than transforming inorganic into organic compounds according to the simulation (Fig.~\ref{b}D), where inorganic compounds seem to require a larger increase of algorithmic information content to reach the complexity of organic compounds, suggesting that inorganic compounds are algorithmically simpler as they are algorithmically more probable, and therefore may occur naturally with much greater ease. Counterintuitively, the results pertaining to organic versus inorganic substances may suggest that organic compounds are much more stable (less reprogrammable) than inorganic compounds in general, and structurally this may be the case, given that the main difference separating the two classes is the stability provided by carbon atoms that are the building blocks of organic matter.

\begin{table}[ht]
    \centering
{\footnotesize
\begin{tabular}{l|r|l|r|l|r}
\textbf{Class}&\textbf{Count}&\textbf{ Class}&\textbf{Count}&\textbf{Class}&\textbf{Count}\\
\hline
  Acids&537   &  Iodinated&774   &  Drugs&4630   \\
  Alcohols&1190   &  Ketones&865   &  Fluorinated&3470   \\
  Aldehydes&723   &  Monomers&994   &  Halogenated&11223   \\
  Amides&531   &  Nitriles&1100   &  Heterocyclic&10964   \\
  Amines&1916   &  Pyridines&1435   &  Inorganic&3605   \\
  Amino Acid Derivatives&799   &  Pyrimidines&558   &  Liquids&10474   \\
  Carboxylic Acids&1145   &  Aromatic&18567   &  Organic&27969   \\
  Esters&1409   &  Biomolecules&6159   &  Organometallic&2315   \\
  Ethers&619   &  Brominated&2620   &  Salts&4718   \\
  Heavy Molecules&1060   &  Chiral&5915   &  Solids&22936   \\
  Hydrocarbons&1499   &  Chlorinated&5680   & \text{} \\
 \hline
\end{tabular}
}
    \caption{Number of elements per compound class used in the experiments. A total of 158\,399 extracted from ChemicalData[] in the Wolfram Language, the sources relied upon being provided in the documentation.}
    \label{table1}
\end{table}

Figure.~\ref{b}(A) illustrates the finding that the class of heavy metals is the least complex according to their algorithmic complexity estimation by BDM. The reason is that most of them tend to be very simple and small while possessing properties that endow them with stronger covalent bonds. In contrast, pyrimidines, for example, which comprise the basis of DNA and RNA nucleotides, are the most complex (together with purines they form
the other nucleotides among the highest complexity heterocylic aromatic organic compounds (see Fig.~\ref{b}(A))). The results show that taking the highest algorithmically complex compounds would profile all pyrimidines (558) with high accuracy, of all the other compounds considered in this database (44\,089), and likewise the lowest complexity retrieves all inorganic compounds followed by heavy and iodinated molecules. The total number of compounds per class can be found in Table~\ref{table1}.

It is also of interest to note that classes of compounds whose algorithmic complexity estimation median values are close to each other are causally related. For example, acids and alcohols appear to have very similar algorithmic information content, notwithstanding the fact that alcohol structures are much larger in size than acids. Their structure and causal origin can be regarded as similar as they can be derived from each other with their main structure unchanged. However, esters are difficult to reduce to ethers, as they decompose to yield alcohols via decomposition of the intermediate hemiacetals even when esters can be reduced to ethers~\cite{yato}.  It is therefore interesting to emulate the evolution of these networks through all possible chemical trajectories and see how algorithmically easy or difficult it is for them to become other compounds favouring certain properties, and how (un)stable they may be in the face of perturbations, independently of thermodynamics, while being ultimately related (and we will devise some tests in this regard). We use a measure of sophistication based on logical depth and a measure of reprogrammability gauging the susceptibility of a chemical compound to being converted into some other, more random or simpler, chemical compound.

\begin{figure}[ht]
  \centering
\includegraphics[width=8cm]{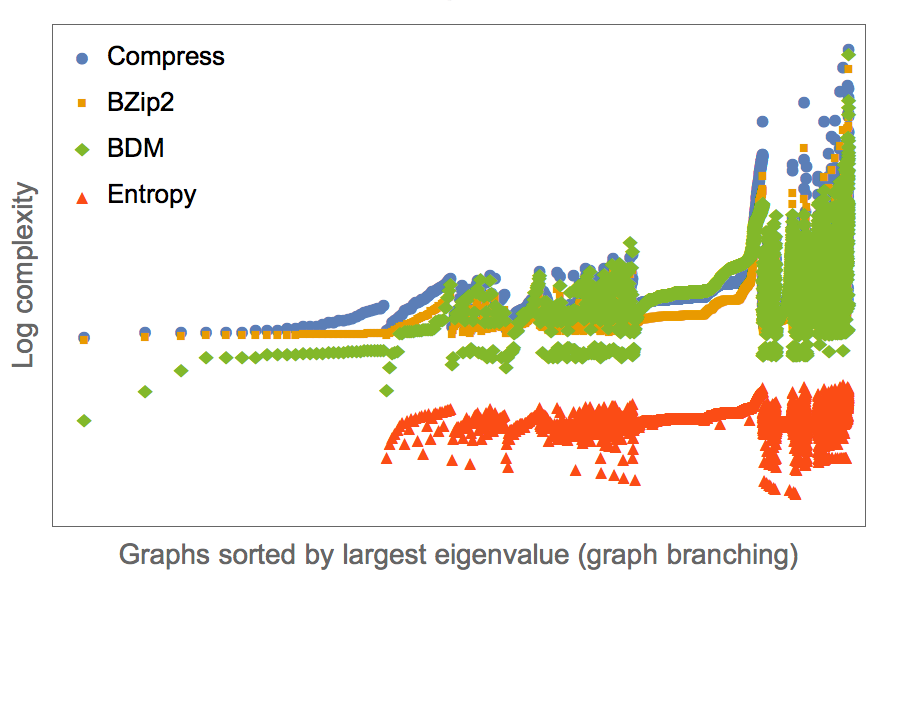}\vspace{-1cm}
   \caption{Graph branching is one of the most important measures proposed in molecular complexity. Here it is shown how algorithmic-based measures such as lossless compression (Compress and Bzip2) are correlated with graph spectra,  which in turn capture graph branching~\cite{randic}. The correlation is stronger than that obtained when using only Shannon entropy. All data come from the Wolfram Language database retrieved by the ChemicalData[] function}
\label{eigenvalues}
\end{figure}

Statistical overlapping of complexity estimations is to be expected, given that classes are not distinct. Aspirin is a drug, but it is also classified as a biomolecule, organic, solid and aromatic. Many classes also share elements with chirality properties. However, a significant divergence between contrasting classes such as those not closely causally related is to be expected, as measured by the algorithmic probability of a compound being in the causal path of another class of compounds (i.e. there being no simple chemical/thermodynamic process---natural or artificial---to convert most elements from one into another class) such as those of an organic nature (including, for example, elements under organic, biomolecules, heterocyclic, pyridines, pyrimidines and monomers) into those of a more inorganic nature (including, for example, the category inorganic itself and heavy molecules).
Other features driving the complexity estimation are properties such as are found in heterocyclic compounds, compounds containing atoms from at least two different elements as members of its rings, thus being structurally richer on average. Nucleic acids and most drugs  are also high in algorithmic complexity and have large variance values, indicating that designed compounds tend to be wide-ranging in nature while inclining towards complexity. Another contrasting/disjoint pair of classes is liquids versus solids, with statistically different complexity values. It is interesting to note that chemical structure networks of liquid compounds are significantly less complex than those of solids, and that inorganic compounds are the least complex, while drugs are the most complex and feature a high incidence of human-made synthetic compounds.

\begin{figure}[h]
  \centering
\includegraphics[width=10.8cm]{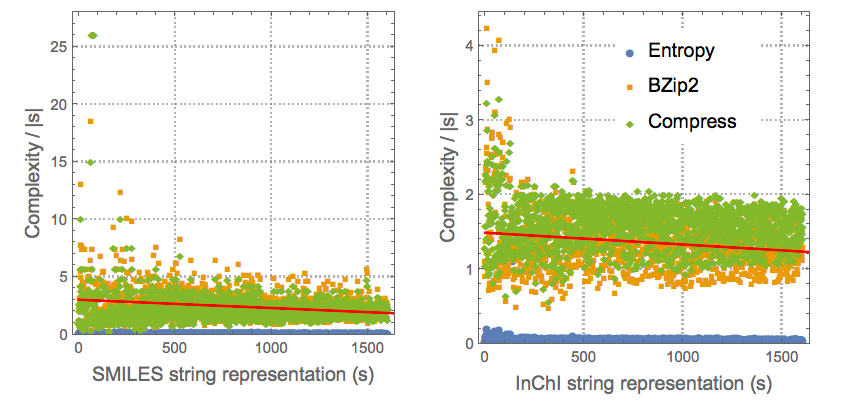}
   \caption{SMILES and InChI molecular strings correlate with complexity even when normalized by size, as would be expected since both notations capture very important properties of the compounds. InChI has a greater correlation statistic, as would also be expected from the fact that the sequence notation was designed to capture more information about the molecular compounds it represents.}
\label{strings}
\end{figure}

\begin{figure}[h]
  \centering
\includegraphics[width=12.5cm]{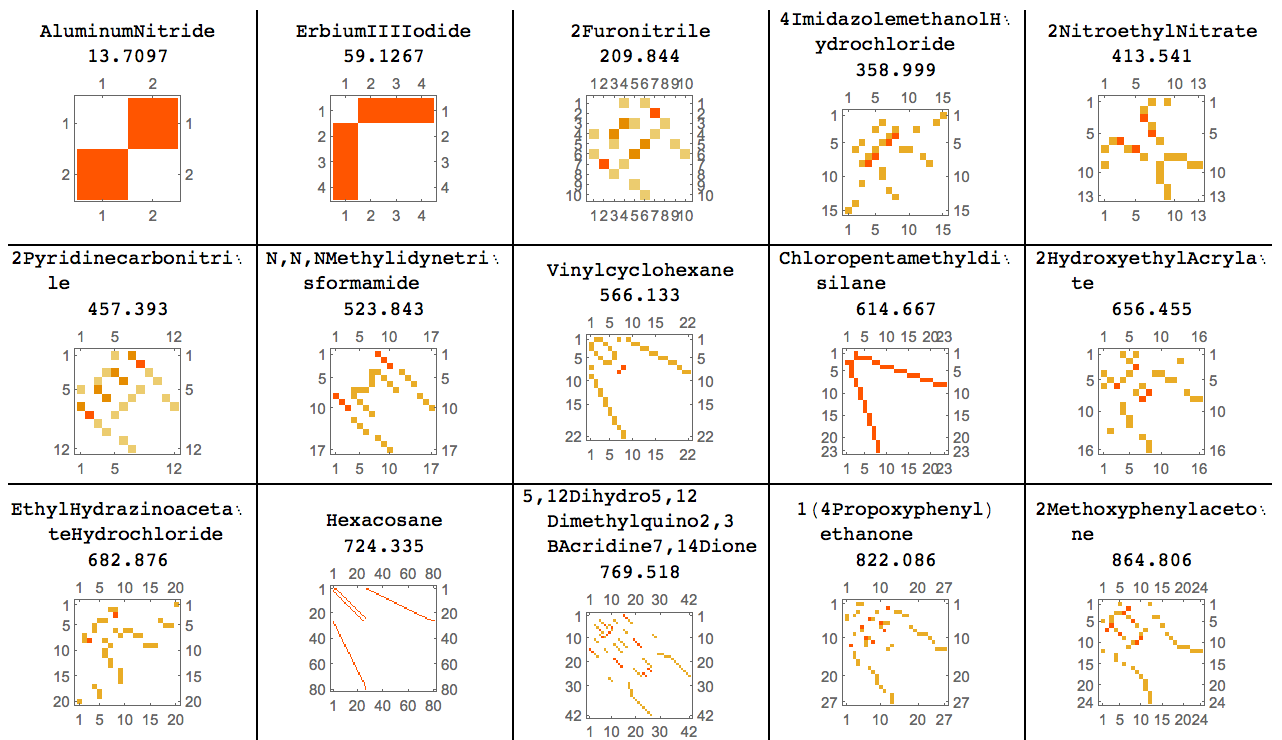}
   \caption{Random sample of chemical structures sorted by algorithmic complexity: BDM sorts molecular graphs by their adjacency matrix, not by their size but their shape. As shown in Fig.~\ref{Memberships} this makes for biases reflected in the class memberships of the molecular compounds, mainly between organic and inorganic, and between solids, liquids and solvents.}
\label{MoleculesSortedByK}
\end{figure}

\subsection{Molecular complexity versus compound properties}

The simplified molecular-input line-entry standards SMILES and InChI are specifications in the form of a line notation for describing the structure of chemical species using short ASCII strings. SMILES is a string obtained by printing the symbol nodes encountered in a depth-first tree traversal of a chemical graph. The chemical graph is first trimmed to remove hydrogen atoms, and cycles are broken to turn it into a spanning tree. Hence one should find some correlations with properties of the represented molecular graph. InChI contains all the information contained in a SMILES description and more atomic information such as bond connectivity, tautomeric information, isotope information, stereochemistry, and electronic charge. Figure.~\ref{strings} reports the correlation found between these notations and complexity by several complexity indexes and Fig.~\ref{eigenvalues} describing the correlation between branching (a common index of order parameter in molecular complexity) and measure of statistical and algorithmic complexity.

\begin{figure}[htbp!]
  \centering
\includegraphics[width=8.5cm]{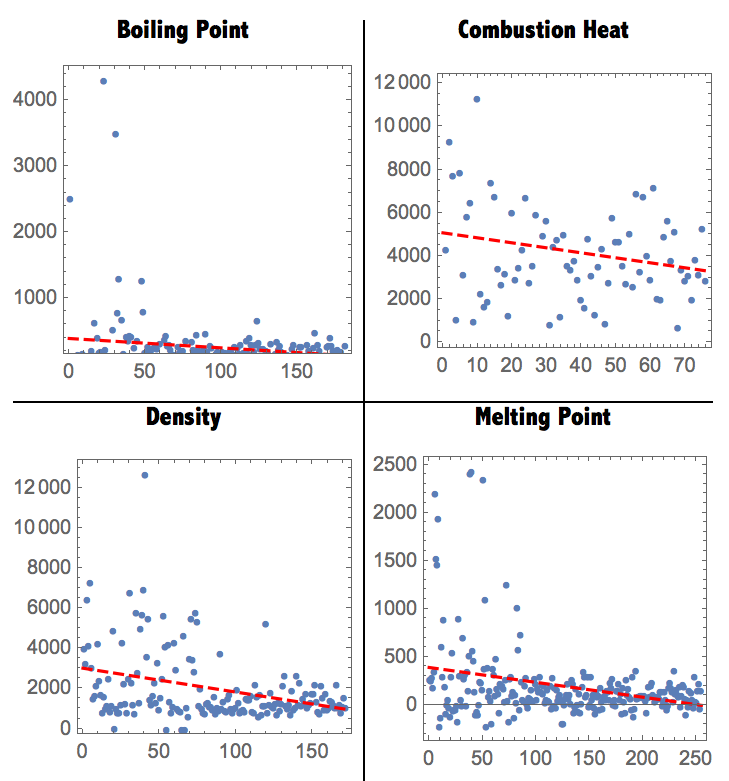}
   \caption{Four physical properties found to be slightly correlated to the graph algorithmic complexity of a large set of molecular networks (built from contact maps extracted from atomic data in ChemicalData[] in the Wolfram Language coming from public chemical data banks) according to their BDM values with statistics (Pearson): -0.16, -0.20, -0.34 and -0.3 and $p$ values all $< 0.05$ except for combustion heat at $p=0.07$. Fitting lines (red) were found by Least square: $x^2 (0.0133001 \text{${}^{\circ}$C})+x (-3.85595 \text{${}^{\circ}$C})+460.328 \text{${}^{\circ}$C}$, $x^2 \left(0.450617 \text{kJ}/\text{mol}\right)+x \left(-57.8591 \text{kJ}/\text{mol}\right)+5521.25 \text{kJ}/\text{mol}$, $x^2 \left(0.0595914 \text{kg}/\text{m}^3\right)+x \left(-22.2082 \text{kg}/\text{m}^3\right)+3306.75 \text{kg}/\text{m}^3$ and $x^2 (0.0144063 \text{${}^{\circ}$C})+x (-5.23141 \text{${}^{\circ}$C})+546.651 \text{${}^{\circ}$C}$. While the correlation values are weak, in all these cases no element with high values for each property was found to also have high BDM. In other words, all elements with high values for each property also had very low algorithmic probability estimations and therefore high algorithmic complexity, thereby pinpointing elements with high values for these physical properties.}
\label{KPhysicalProperties}
\end{figure}
\begin{figure}[htbp!]
  \centering
\includegraphics[width=8.5cm]{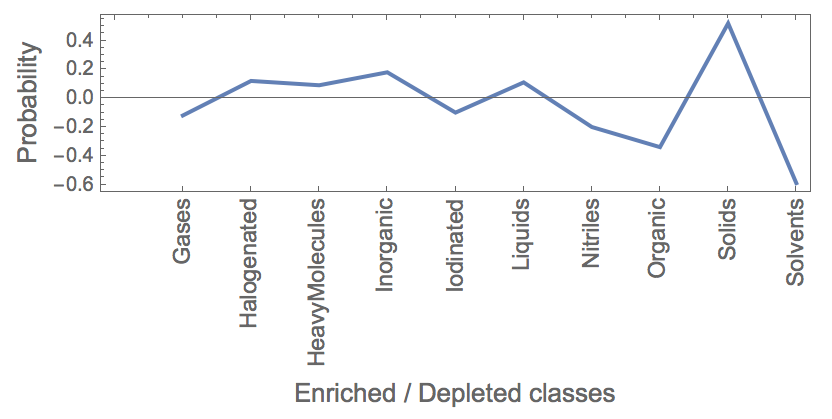}\\

\bigskip

\includegraphics[width=8.2cm]{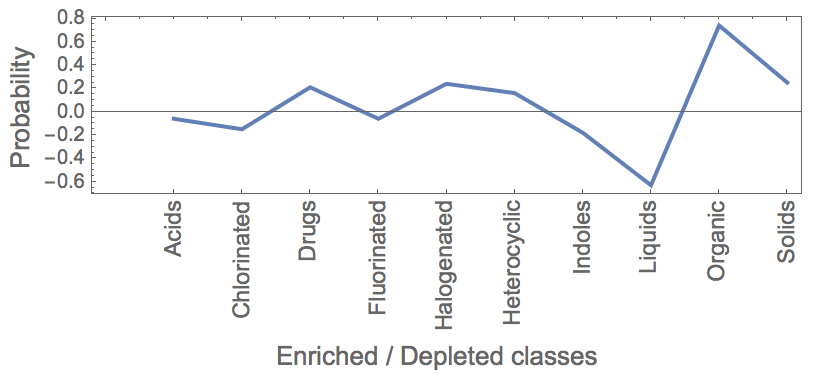}
   \caption{Top enriched and depleted classes separated by estimations of graph algorithmic complexity (approximated by CTM/BDM v. randomized membership) out of a random sample of 100 complex chemicals sorted by class. The probability on the $y$-axis is given by the Spearman correlation betweeh complexity values and class membership. Liquids, solids, organics and solvents are the best separated when enriched by algorithmic information content, i.e. with generating mechanisms with similar computer program lengths when the underlying networks are explained causally (as generated by 2-dimensional computer programs). Data source: Wolfram Language database from the ChemicalData[] function. }
\label{Memberships}
\end{figure}

The results reported in Fig.~\ref{MoleculesSortedByK} (Appendix) are interesting because one can think of algorithmic complexity as a method sorting by size of the underlying generating mechanism causing each structure, and it would be expected to find compounds that behave or produce similar states to be generated by similar mechanisms. The results reported together with the class enrichment and depletion analysis shown in Fig.~\ref{Memberships}(Appendix) are interesting because, as has been suggested, in the case of organic molecules, the lower the information content the fewer the possibilities for different interactions with other molecular compounds. Fig.~\ref{KPhysicalProperties} illustrates how the algorithmic complexity approximated by BDM correlates with some physical properties of the molecular compounds. This is interesting because it suggests-- if the correlation is actually correct-- that some information about properties that are global properties, such as temperature, is in the local structure of the molecule, which is not surprising if one recalls that how rigid or charged a particle is may have an impact on its dynamic interactions with other molecular compounds.

\section{Conclusions}

We have identified and illustrated interesting research avenues that a causal interventional calculus based on algorithmic probability/complexity (as the study of a system's changes in algorithmic information) can bring to the discussion of molecular complexity and chemoinformatics, in particular to chemical structure networks. Indeed, usually a drug has a backbone from high throughput screening leading hits and intensive modification of the chemical structure is done to make it drug like, providing improved stability, solubility etc.
 
We have found that this algorithmic approach suggests similarities for graphs/networks that may be explained by common generative mechanisms, suggesting an algorithmic likelihood of causal transformations from candidate models found by algorithmic probability. We found that the method separates distinct classes of chemical compounds, both by estimations of the algorithmic complexity of chemical structures and for causal sensitivity based on a measure of agnostic (no physical properties being involved) reprogramability. The experiments with statins whose similar effects to aspirin are an open question, suggests a measure of similarity and of interactions/toxicity that should be further tested. We have also shown that the measures show various degrees of correlation with some chemical compound's physical properties.

Our approach effectively introduces a new dimension in the study of information-theoretic properties and algorithmic transformations of compounds, and further explorations and generalizations to more general compound networks (where several compounds are connected), binding and reaction networks should be investigated.

\section*{Acknowledgements}

H.Z. was supported by the Swedish Research Council (Vetenskapsr\r{a}det) grant No. 2015-05299.

\newpage

\section*{Appendix}

\end{document}